\documentclass[]{aastex701}

\usepackage{natbib}

\shorttitle{Surface Interactions}
\shortauthors{Peterson et al.}

\begin{document}
\nolinenumbers

\title{Surface Interactions in Photon Monte Carlo Simulations}

\author{J.~R.~Peterson}
\affiliation{Department of Physics and Astronomy, Purdue University, West Lafayette, IN 47907, USA}
\email{peters11@purdue.edu}

\author{D.~Valls-Gabaud}
\affiliation{LUX, CNRS UMR 8262, Observatoire de Paris, PSL, 61 Avenue de l'Observatoire, F-75014 Paris, France}
\email{david.valls-gabaud@obspm.fr}

\author{A.~Dutta}
\affiliation{Department of Physics and Astronomy, Purdue University, West Lafayette, IN 47907, USA}
\affiliation{National Center for Nuclear Research, ul. Pasteura 7, 02-093 Warsaw, Poland}
\email{dutta26@purdue.edu}

\author{C.~Kim}
\affiliation{School of Space Research and Institute of Natural Science, Kyung Hee University, 1732 Deogyeong-daero, Giheung-gu, Yongin-si, Gyunggi-do, Republic of Korea, 17104}
\email{changgonkim@khu.ac.kr}

\author{G.~Sembroski}
\affiliation{Department of Physics and Astronomy, Purdue University, West Lafayette, IN 47907, USA}
\email{sembrosk@purdue.edu}

\correspondingauthor{John~R.~Peterson}

\begin{abstract}

  We implement a comprehensive simulation of photon surface interactions using a Monte Carlo approach.  This is effective in simulating the interaction of light with telescope mirrors and lenses.  We use a full electromagnetic solution to simulate the wavelength and angular dependence at surfaces.  This includes bare interfaces, monolayer interfaces, protected layer coatings, and multilayer coatings.  We handle special cases when multilayer data is incomplete or when there is photo-conversion in the interface as with sensors.  We implement interactions with surface micro-roughness and predict the corresponding angular distribution using a Monte Carlo implementation of the Harvey-Shack scatter theory for a microroughness power spectrum.
  Finally, we simulate surface interaction with contamination from dust or condensation using Mie scattering applied efficiently to individual contaminants.  The combination of these implementations can efficiently simulate rough to polished surfaces of arbitrary materials that are fully cleaned or dusty.  The observational consequences includes complex wavelength and spatial-dependent photometric errors, the dominant effect of the wings of point-spread-functions, dust rings, and wavelength and angle-dependent throughput losses.  We find agreement with the point-spread-function wings of WIYN ODI observations of bright stars and properties of dust rings, and demonstrate the ability to disentangle mirror microroughness from the turbulence PSF patterns.  The comprehensive numerical implementation of surface interactions has wide applicability in non-astronomical applications as well.
  
\end{abstract}


\section{Introduction}

Simulations are critical to understanding a physical system and how it relates to complex measurements.  In modern astrophysics, precision measurements in photometry, astrometry, and object shape have increased the need for simulations.  A number of different methods have been used to simulate astronomical observations.   In both high-energy astrophysics (\citealt{peterson2004}; \citealt{peterson2007}; \citealt{andersson2007}; \citealt{davis2012}) and optical astrophysics (\citealt{lane}, \citealt{ellerbroek}; \citealt{lelouarn}; \citealt{britton}; \citealt{jolissaint}; \citealt{bertin}; \citealt{dobke}; \citealt{ackermann2012}; \citealt{peterson2015}, \citealt{rowe2015}, \citealt{arko2022}) codes have been constructed to simulate realistic images or simulate part of the observational process.

In our previous work, we created an \textit{ab initio} physics simulator using a photon Monte Carlo approach to simulate realistic images called the Photon Simulator\footnote{\url{https://www.phosim.org}} \citep[PhoSim;][]{peterson2015}.   This code generates photons from astrophysical sources and follows their interaction with the atmosphere, optics, and sensors of a given observatory.  We have extended the simulation to include the thermal and mechanical deformation of optics (\citealt{peterson2019}), the detailed electrostatic physics of sensors (\citealt{peterson2020}), and the self-consistent simulation of the atmosphere (\citealt{peterson2024}).   Thus, the simulator uses first principle physics to simulate the interaction of photons (and subsequent electrons) as well as includes the relevant physics of the atmosphere, optics, and sensor.

There are a diverse set of applications that can take advantage of a photon simulator with comprehensive physics.   PhoSim has been used to test the design of telescopes (\citealt{xin2015}, \citealt{angeli2016}, \citealt{thomas2016}, \citealt{burke2019a}) and prepare and predict future observations (\citealt{chang2012}, \citealt{chang2013a}, \citealt{chang2013b}, \citealt{bard2013}, \citealt{mandelbaum2014}, \citealt{bard2016},  \citealt{thomas2018}, \citealt{sanchez2020}, \citealt{merlin2023}, \citealt{bretonniere2023}).  PhoSim has also been used for advanced algorithm development (\citealt{meyers2015}, \citealt{li2016}, \citealt{carlsten2018}, \citealt{nie2021a}, \citealt{nie2021b}) including studies to understand physical effects from complex physics (\citealt{walter2015}, \citealt{beamer2015}).  Recently, PhoSim has been used to generate perfect knowledge training sets for machine learning/artificial intelligence (AI) applications (\citealt{burke2019b}, \citealt{yang2023}, \citealt{merz2023}).  This may be the future approach for disentangling complex emergent effects, such as those predicted by PhoSim, by directly inverting measured quantities from a PhoSim-trained AI.

Recent work with PhoSim has focused on both the physical aspects of the objects of the system and refining photon interactions so they are based on first principles physics rather than parameterized models.  For the interactions of light with the optics and sensor, we previously described the raytracing implementation in \cite{peterson2015} for arbitrary perturbed surfaces.   This interface can use elasticity physics from thermal and mechanical deformations in \cite{peterson2019}.   Here we extend those interactions to calculate the electromagnetic boundary conditions for surface interfaces including coatings.  This results in a complex photometric response that depend on the angle, position, and wavelength.

We also describe a method to properly simulate surface microroughness using Harvey-Shack scatter theory with a wavelength and angular dependence that relates to the underlying roughness of the surface.  This allows us direct physical theories based on surface properties rather than use empirical models.  The scattering results in a large point-spread-function (PSF) wing, and is accurate enough to describe real observations.  This can be important in source photometric estimates where modelling the profile as it extends below the background level.  It can also be critical for deblending sources in dense fields.  More generally, this scattering is responsible for a physical surface not behaving as a perfect mirror or lens.  Finally, we also implement the effect of dust and condensation on surfaces.  This also leads to photometric errors and can cause complex background patterns.  All of this work is available publicly with PhoSim (\url{https://www.phosim.org}), and is therefore reproducible.

In this work, we extend the interactions of photons to complex interactions with surfaces.  For a telescope, this would include the surfaces of the mirrors, lenses, and the sensor.  We first describe the implement of electromagnetic solutions to uncoated interfaces, monolayer interfaces, protected layer coatings, and multilayer coatings in \S2.  Then in \S3 we apply Harvey-Shack scatter theory to micro-roughness on surfaces to predict the angular and wavelength distribution of scattered light on surfaces with a specified power spectrum.   In \S4, we apply Mie scattering to debris or condensation on surfaces to predict photon interactions and implement this in an efficient manner to simulate contaminants on large optical surfaces.  In \S5 and \S6, we discuss the integration of these techniques with the overall approach and discuss future work and conclusions.

\section{Interface Reflectivity \& Transmission}

By considering the electromagnetic boundary conditions on the surface interface, the wavelength and angle-dependent reflection, transmission, and absorption properties can be predicted.   This is done by considering the electric and magnetic field directions of the propagating wave which is determined by the polarization and satifying the boundary conditions by considering the normal to the surface interface.  This can then be used to predict the future propagation of the photon.  Below, we describe many situations where the conditions can be evaluated.

\subsection{Bare Interfaces}

We first consider the simple case of a surface dividing two different materials.  Often one medium is air or vacuum and the other is glass, but there are many other situations where two different media meet at a common surface boundary.  The reflection coefficient, $r$, is then entirely determined by the familiar solution of Fresnel (\citealt{fresnel1821}),

\begin{equation}
  r = {\vline \frac{ n_i { \cos(\theta_i) } ^ p - n_o { \cos(\theta_o) } ^p}{ n_i { \cos(\theta_i) } ^ p + n_o { \cos(\theta_o) } ^p}  \vline}^2
  \label{equation1}
  \end{equation}

\noindent
where $n_i$ and $n_o$ are the (possibly complex) index of refraction of the incoming ($i$) and outgoing ($o$) media, and $p$ is the polarization (+1 for s-polarization and -1 for p-polarization).   The outgoing angle is deduced from Snell's law (\citealt{sahl984}, \citealt{rashed1993}) as

\begin{equation}
   n_i \sin(\theta_i) = n_o \sin(\theta_o)
  \label{equation2}
  \end{equation}

\noindent
using the real part of the indices of reflection.  For a 3-dimensional raytrace the angles are determined by taking the scalar product with the surface normal at the intercept location and the intercept location has to be determined iteratively (\citealt{peterson2015}).  In this simple case, the transmission coefficient is determined by $1-r$.  The interaction of light is then entirely determined by both the wavelength-dependent complex index of refraction and the polarization.

\subsection{Semi-Infinite Coatings}

A special case of the bare interface solution is for a surface that is coated with a sufficiently thick coating.  This occurs when the product of imaginary part of the index of refraction with the thickness is much larger than the photon wavelength ($\mbox{Im} n \times  z \gg \lambda$) .  In this case, the coating itself behaves as the outgoing media described above, but the transmission instead can be interpreted entirely as absorption since we effectively lose the photon.  If the coating is sufficiently thick, then the possible reflection from the substrate can be ignored.  Therefore, for sufficiently thick (i.e. semi-infinite) coatings we can use the same equations interpreted differently as the bare interfaces.

\subsection{Monolayer Coatings}

Most surfaces on telescope optics and sensors are coated in order to either maximize or minimize the reflections from the Fresnel equations.   In this case, the index of refraction of the coating, $n_m$, and the thickness, $z$, need to be considered.  The solution can be solved by considering the set of boundary conditions for the two surfaces on either side of the coating can be written following \cite{orfanidis} in terms of Fresnel coefficients as

\begin{equation}
  \rho_1 = \frac{ n_i { \cos(\theta_i) } ^ p - n_m { \cos(\theta_m) } ^p}{ n_i { \cos(\theta_i) } ^ p + n_m { \cos(\theta_m) } ^p}
  \label{equation3}
  \end{equation}

\begin{equation}
  \rho_2 = \frac{ n_m { \cos(\theta_m) } ^ p - n_o { \cos(\theta_o) } ^p}{ n_m { \cos(\theta_m) } ^ p + n_o { \cos(\theta_o) } ^p}
  \label{equation4}
  \end{equation}

\noindent
where the $i$, $m$, $o$ subscripts refer to the incoming, coating, and outgoing complex index of refractions as well as the propagation angle.  The angles can be calculated from Snell's law as

\begin{equation}
  n_i \sin(\theta_i) = n_o \sin(\theta_o)
  \label{equation5}
  \end{equation}

\begin{equation}
  n_m \sin(\theta_m) = n_o \sin(\theta_o)
  \label{equation6}
  \end{equation}

\noindent
The reflectivity can then be calculated by determining

\begin{equation}
  r = {\vline \frac{ \rho_1 + \rho_2 e^{-2 i \delta} } { 1 + \rho_1 \rho_2  e^{-2 i \delta}} \vline}^2
  \label{equation7}
  \end{equation}

\noindent
where $\delta$ is given by

\begin{equation}
  \delta = \frac{2 \pi}{\lambda} z n_m \sqrt{ 1 - \left( \frac{ \sin(\theta_i) } {n_m} \right)^2 }
  \label{equation8}
  \end{equation}

\noindent
It can often be assumed that the coating is lossless if it is sufficiently thin, so that the transmission is simply given by $1-r$.  If the coating has significant attenuation, then the transmission would be given by $1 - r {\vline e^{-2 i \delta} \vline}^2$.

\subsection{Monolayer on Semi-Infinite Coatings}

Similar to the semi-infinite coating discussed above, a monolayer coating on top of a sufficiently thick (i.e. semi-infinite) coating can be treated in the same manner.  In this case, the transmission past the second interface is treated as absorption and the photon is lost.

\subsection{Multilayer Coatings}

For more than one layer, we can use the recursion relations that satisfy the boundary conditions as in \cite{orfanidis}.  At each interface, we compute

\begin{equation}
  \rho_i = \frac{ n_i { \cos(\theta_i) } ^ p - n_{i+1} { \cos(\theta_{i+1}) } ^p}{ n_i { \cos(\theta_i) } ^ p + n_{i+1} { \cos(\theta_{i+1}) } ^p}
  \label{equation9}
  \end{equation}

\noindent
and then evaluate

\begin{equation}
  \Gamma_i = \frac{ \rho_i + \Gamma_{i+1} e^{-2 i \delta_i}}{1+ \rho_i \Gamma_{i+1} e^{-2 i \delta_i}}
  \label{equation10}
  \end{equation}

\noindent
where $\delta_i$ is the same as in the monolayer case.  For $N$ layers, then the recursion relations are evaluated in reverse where $\Gamma_N = \rho_N$.  Then the reflectivity is calculated by determining $r = { \vline \Gamma_i \vline}^2$ and transmission assuming lossless media is $1-r$.  We have applied this in the case of a sensor where a monolayer is placed on Silicon forming the bi-layer between vacuum layers resulting in three surfaces.

Many layers are quite common in astronomy with the use of interference filters.  However, in many cases the multilayer prescription is often proprietary to the optic manufacturer.  In this case, the reflection and transmission as a function of wavelength and angle are often provided.   We then use the numerical table as $T(\lambda, \theta)$ and $R(\lambda, \theta)$ in a four column format and interpolate for a given photons as in \cite{peterson2015}.  We can use these tables as the starting point, but go beyond this information for two reasons.  One is that the angular dependence is often coarse or limited to the pure normal incidence case.   The second is that it does not allow for the consideration of coatings with significant non-uniformity variation as we describe below.

To address this, we have found a useful approximation for use when the full multilayer prescription for layer materials and thicknesses  is not known.  First, we find an approximation when angular distribution is unknown where we might know $R(\lambda, \theta)$ and can find $R(\lambda, \theta^{\prime})$ by looking up $R(\lambda^{\prime}, \theta)$ where $\lambda$, $\lambda^{\prime}$, $\theta$, and $\theta^{\prime}$ is defined by

\begin{equation}
   \lambda^{\prime} \sqrt{ 1 - \left( \frac{ \sin(\theta_i^{\prime}) } {n_m} \right)^2 } = \lambda \sqrt{ 1 - \left( \frac{ \sin(\theta_i) } {n_m} \right)^2 } 
  \label{equation12}
  \end{equation}

\noindent
Similarly, for non-uniformity, we can shift the distribution according to $\lambda^{\prime} = \lambda z^{\prime} / z$.  Both of these approximations are accurate because of the role of $\delta_i$ in the above equations.

We assessed the accuracy of this approximation by using an example with the Rubin observatory filter transmission curves that have a complex multi-layer coating.  Here we have an angle and wavelength-dependent transmission and reflection table, but not the full multi-layer prescription.  Then we pretend we do not know the transmission curve for a particular angle, and interpolate using the above approximation instead.  For the six filters, we find a filter curve with similar features, an average accuracy of 0.3$\%$, and a value of $n_m=1.7$ even when we interpolate by 10 degrees.  The more information that is available about the multi-layer prescription makes this additional step unnecessary, but 0.3$\%$ is generally several times less than typical the non-uniformity that appears to be present in constructed filters as we discuss below.

\subsection{Electron Conversion}

A special case of the formalism above is to consider what happens when absorption occurs, which is important in the case of a sensor where the photo-electron will be tracked.  First, we can consider the probability implied by the reflection and transmission of each surface.  If absorption occurs, then the simulation shifts to propagating an electron at the start of the surface.  We then use the mean free path set to be equal to $\lambda / \left( 4 \pi \mbox{Im} n \right) $ where $\mbox{Im} n$ is the imaginary part of the index of refraction.   Then the electron is propagated along its path by a distance following an exponential distribution with the calculated mean free path.  This works well when applied to a sensor because it correctly predicts the photon conversion rate and produces detailed interference effects like fringing (\citealt{peterson2015}, \citealt{peterson2020}).

\subsection{Coating Examples}

\begin{figure}[htb]
\begin{center}
\includegraphics[width=0.99\columnwidth]{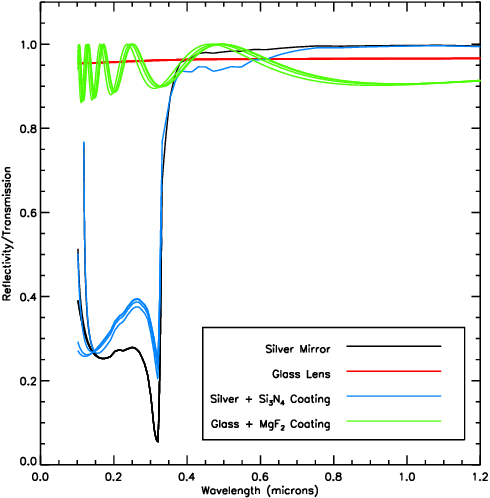}
\end{center}
\caption{\label{fig:label1} Reflectivity or transmission as a function of wavelength.  A silver-coated mirror and a fused Silica lens are simulated.  We then add a Silicon Nitride and Magnesium Fluoride coating, respectively.  For each case, simulations are performed at  0, 10, 20, and 30 degrees the corresponding curves are plotted.}
\end{figure}

In Figure~\ref{fig:label1}, we simulate various reflectivity functions.   We demonstrate the reflectivity of a Silver coated mirror as a function of wavelength at 0, 10, 20, and 30 degree off-axis incidence angles as an example of a semi-infinite coating using optical constants from \cite{johnson1972}.  We then add a 0.083 micron Silicon Nitride coating that enhances the reflectivity at 0.3 microns to show the implementation of the monolayer with a semi-infinite coating with optical constants from \cite{philipp1973}.  Similarly, we show the transmission of a fused Silica lens as a function of wavelength at 0, 10, 20, and 30 degree incidence angles to show a bare intereface using optical constants of \cite{malitson1965}.  We then add a monolayer of Magnesium Fluoride of 0.18 microns that enhances the transmission at 0.5 microns to also test the monolayer functionality with optical constants from \cite{li1980}.

\subsection{Coating Non-uniformity}

The reflectivity formalism above is most important when considering the complicated case of non-uniformity of individual layers of any of the scenarios described above.  This turns out to be rather important and has been studied by \cite{yao2002}.   They demonstrated that a significant source of photometric error could exceed 0.05 magnitudes for systems with interference coatings, and made detailed maps of the spatial non-uniformities of coated optics.  In a Monte Carlo approach, complex spatial patterns can be considered for any layer.   We implement this by using Zernike polynomials for every circular surface (e.g. lenses) and Chebyshev polynomials for any rectangular surface (e.g. sensors).  The input is therefore the layer thickness (zeroth order component) and higher-order coefficients in units of thickness.

\begin{figure}[htb]
\begin{center}
\includegraphics[width=0.99\columnwidth]{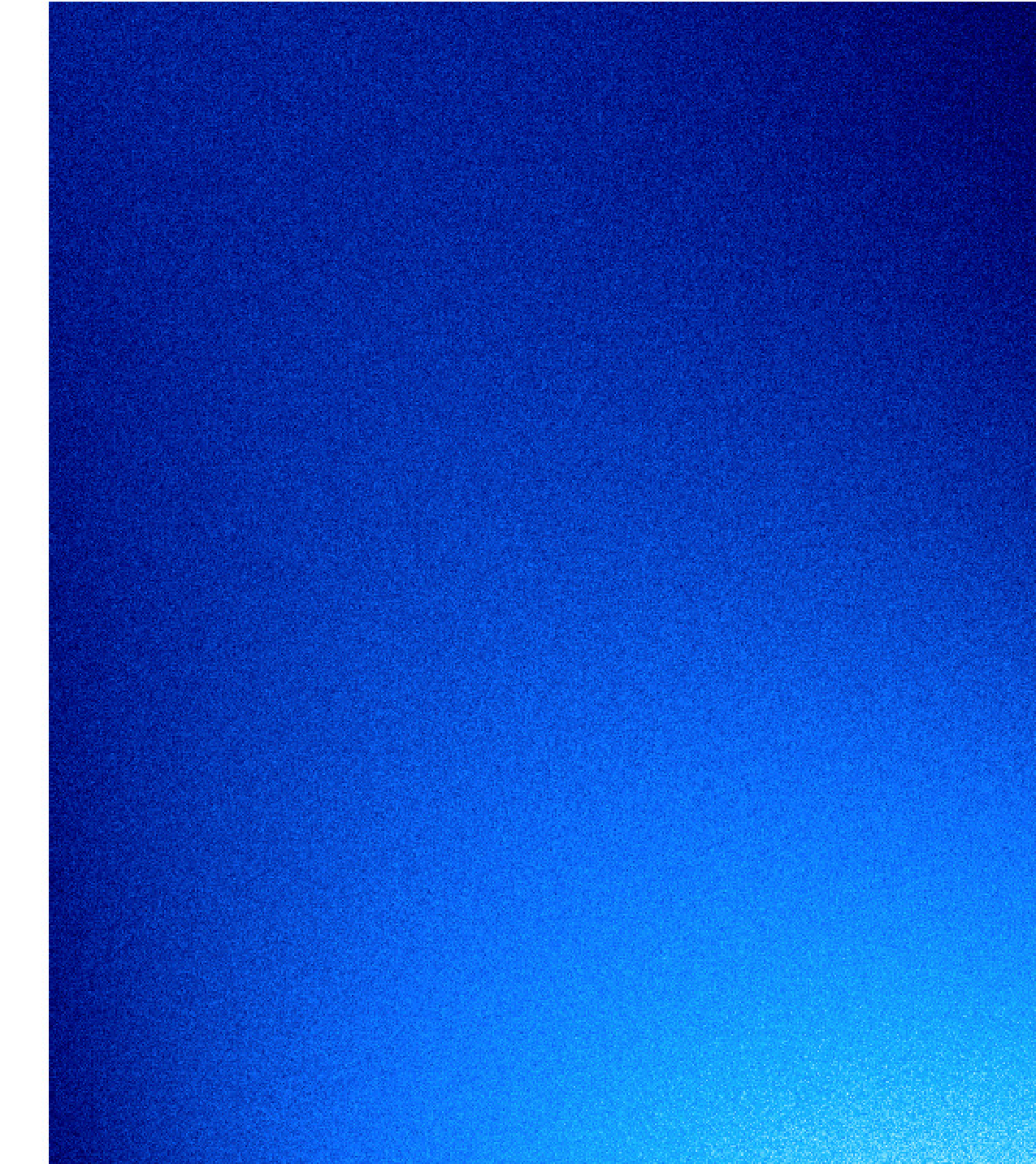}
\end{center}
\caption{\label{fig:label2} Simulation of the background across a chip of a generic 1 m telescope with a $r$-band filter that has a non-uniform coating.  The amplitude of the non-uniformity is set to 10\% and the corresponding variation across the chip is also about 10\%.}
\end{figure}

Then, we use the approximation that the nominal eflectivity $R(\lambda, \theta)$ at a given location is modified by $R(\lambda z(x,y)/z_0, \theta)$.  This then uses the same approximations as with multilayers and avoids having to evaluate the calculation for every photon, since $R(\lambda, \theta)$ are already stored on a grid.  In this way, we capture the wavelength-dependence, angle-dependence, and spatial non-uniformity.  An example of this is shown in Figure~\ref{fig:label2}.  There we simulate a 1 m generic telescope with a filter that has a total of 10\% non-uniformity of a $r$-band filter function that we capture in the first 10 Zernike terms with random amplitudes.  The overall variation of the photometric response is quite significant for a interference filter, since it can be of order the non-uniformity.  With anti-reflective coatings, the variation may be smaller since the transmission function is smooth and not designed to reject photons.  This is in agreement with the analysis of \cite{yao2002}.

\section{Microroughness}

An important consideration of the interaction of light with surfaces is the small-scale microroughness patterns.  This results in significant scattering to large angles, and is the reason surfaces are not perfectly reflective.  Even in astronomy with careful weeks-long process of polishing of a mirror, a significant amount of scattering still occurs.

\subsection{Harvey-Shack ABC scattering}

The microroughness often occurs with a scale smaller than the photon wavelength, so a full diffractive approach is needed.  This is unlike the surface figure and large-scale perturbations that can be treated with geometric raytracing as in \cite{peterson2015}.  The surface scattering of a rough surface has been treated in Rayleigh-Rice theory (\citealt{rice1951}, \citealt{church1979}) for sufficiently smooth surfaces and in Beckmann-Kirchoff theory (\citealt{beckmann1963}) for moderate roughness at small angles.  A method that comprehensively predicts the scattering of light off surfaces of moderate roughness at all angles is Harvey-Shack theory (\citealt{harvey1976}, \citealt{harvey1989}, \citealt{vernold1998}, \citealt{harvey2007}).  Since our problem involves large angle scattering off moderate surfaces we will apply Harvey-Shack theory.  Note that the scattering primarily concerns microroughness of mirror or reflective surfaces since in transmissive lens surfaces the phase shift induced by the microroughness approximately cancels.

To apply the scattering theory to an arbitrary simple surface, a description of the roughness pattern must be developed.  From metrology studies of real surfaces a fractal representation describing the spatial power spectrum works well (\citealt{church1988}, \citealt{hasan1995}).  This is characterized by a power-law spatial frequency spectrum, but this formulation diverges at low frequency.  To avoid the divergence the so-called ABC model (\citealt{church1991}, \citealt{elson1993}, \citealt{stover1995}, \citealt{dittman2006}) flattens the spectrum at low spatial frequency.  The application of Harvey-Shack scattering to the ABC model is comprehensively described in \cite{krywonos2011}.  Following \cite{krywonos2011} the two-dimensional spatial power spectrum of a surface can be written as

\begin{equation}
  PSD(f) = K \frac{AB}{{\left[ 1 + {\left( B f \right)}^2 \right]}^{\frac{C+1}{2}}}
  \label{equation13}
  \end{equation}

\noindent
where $A$, $B$, and $C$ are the parameters of the ABC model that describe the normalization, cutoff inverse frequency, and slope, respectively.   $f$ is the 2-d spatial frequency, and $K$ is defined by

\begin{equation}
K =  \frac{1}{2 \sqrt{\pi}} \frac{ \Gamma \left( \frac{C+1}{2} \right) }{ \Gamma \left( \frac{C}{2} \right) }
  \label{equation14}
  \end{equation}

\noindent
This representation  implicitly assumes isotropic roughness.   The expression can be further integrated to give

\begin{equation}
  \sigma^2 = \frac{2 \pi K A B}{ (C-1) B^2}
  \label{equation15}
  \end{equation}

\noindent
where $\sigma$ is then a measure of the total roughness.  Note that this will be larger from the measured roughness which is usually band-limited (see \cite{pfisterer2012}).  From this formulation, \cite{krywonos2011} derived the bidirectional reflectance function (BRDF) as

\begin{equation}
  S = \frac{ \frac{1}{2} KAB} { \left[ 1 + \left(  \frac{B \alpha}{2 \pi} \right)^2 \right]^{\frac{(C+1)}{2}} } \beta^2 \left( \frac{ 4 \pi \cos{\theta_i} }{ \lambda^2} \right)
  \label{equation16}
  \end{equation}

  \noindent
  where

  \begin{equation}
    \beta = \frac{2 \pi}{\lambda}  \left( \cos{\theta} + \cos{\theta_i} \right)
    \label{equation17}
    \end{equation}

 \noindent
and

 \begin{equation}
   \alpha = \frac{2 \pi}{\lambda}  \sqrt{ {   \left( \sin{\theta} \cos{\phi} - \sin{\theta_i} \right) }^2 + \left( \sin{\theta} \sin{\phi} \right)^2 }
   \label{equation18}
   \end{equation}

 \noindent
 where $\theta$ and $\phi$ are the polar angles of the emerging photon.  Here we have defined the normalization in the last factor from \cite{krywonos2011}, so the total integrated scatter, $T$, follows Rayleigh-Rice theory as

  \begin{equation}
    T = \left (\frac{4 \pi \sigma \cos{\theta_i}}{\lambda} \right)^2
    \label{equation19}
  \end{equation}

  \noindent
 This is the limit of small roughness, and in the next section we extend this to arbitrary roughness.
 
  \subsection{Monte Carlo implementation}

  It is counterintuitive that the implementation of the scattering described above with no significant approximations in a photon Monte Carlo is actually not straight-forward.   In the context of simulating millions of photons per second in PhoSim, evaluating the BRDF analytically for each photon would be computationally prohibitive.   We also found that saving the cumulative BRDF in a two-dimensional table ($\phi, \theta$) that can be evaluated at runtime would not be possible.  This is because in many cases the profile at sub-arcsecond scales is desired around the reflection point as this forms the point spread function wing.  On the other hand, the profile extends over the entire solid angle, so the size of the table in memory would be prohibitive.  It is possible to use a optimized non-linear grid, but we outline a better solution below.

 To implement in a Monte Carlo approach, we first evaluate the scattering fraction according to

 \begin{equation}
   1 - e^{- {\left( \frac{ 4 \pi \sigma \cos{\theta_i}}{\lambda} \right)}^2}
   \label{equation20}
   \end{equation}

   \noindent
   to first decide whether the photon will be scattered or not.  This formulation leads to a scattering fraction of unity in the limit of a rough surface, and presumably this is the correct limit of the Harvey-Shack scattering for arbitrary roughness.   Then, we need to determine the scattering angle from the BRDF above.  To implement an efficient Monte Carlo approach we note that the BRDF is mostly a function of $\alpha$ with a minor slowly varying function of $\beta$.   Therefore, we can use a Monte Carlo separation of variables-type approach.   We first can calculate the probability distribution, $P(\alpha)$, and determine the cumulative distribution at a large number of points since it is a one-dimensional distribution.

   \begin{equation}
     P = \alpha   \left[ 1 + \left( \frac{B \alpha}{\lambda} \right)^2 \right]^{-\frac{C+1}{2}}
     \label{equation21}
     \end{equation}

\noindent
Furthermore, since we want to simulate photons at many wavelength simultaneously, it is useful to formulate the function in terms of $\alpha^{\prime} = B \alpha / \lambda$.  We first choose $\alpha^{\prime}$ by choosing a random number and matching the cumulative distribution.  Then the value of $\alpha$ is determined depending on $B$ and the wavelength, $\lambda$.

After determining $\alpha$, an azimuthal angle, $\psi$, is uniformly chosen.  Then, the polar angles are chosen by

\begin{equation}
  \phi = \arctan{ \frac{\alpha \sin{\psi}}{ \sin{\theta_i} + \alpha \cos{\psi}} }
  \label{equation22}
\end{equation}

\begin{equation}
  \theta = \sqrt{ {\left( \sin{\theta_i} + \alpha \cos{\psi} \right) }^2 + { \left( \alpha \sin{\psi} \right) }^2}
  \label{equation23}
  \end{equation}
  
\noindent
Finally, in order to evaluate the slowly varying function which is a function of $\beta$, we first determine $\theta$ by

\begin{equation}
  \cos{\theta} = \sqrt{ \cos{\theta_i}^2 - \alpha^2 - 2 \alpha \sqrt{1-{\cos{\theta_i}}^2} \cos{\psi} }
  \label{equation24}
  \end{equation}

\begin{figure}[htb]
\begin{center}
\includegraphics[width=0.99\columnwidth]{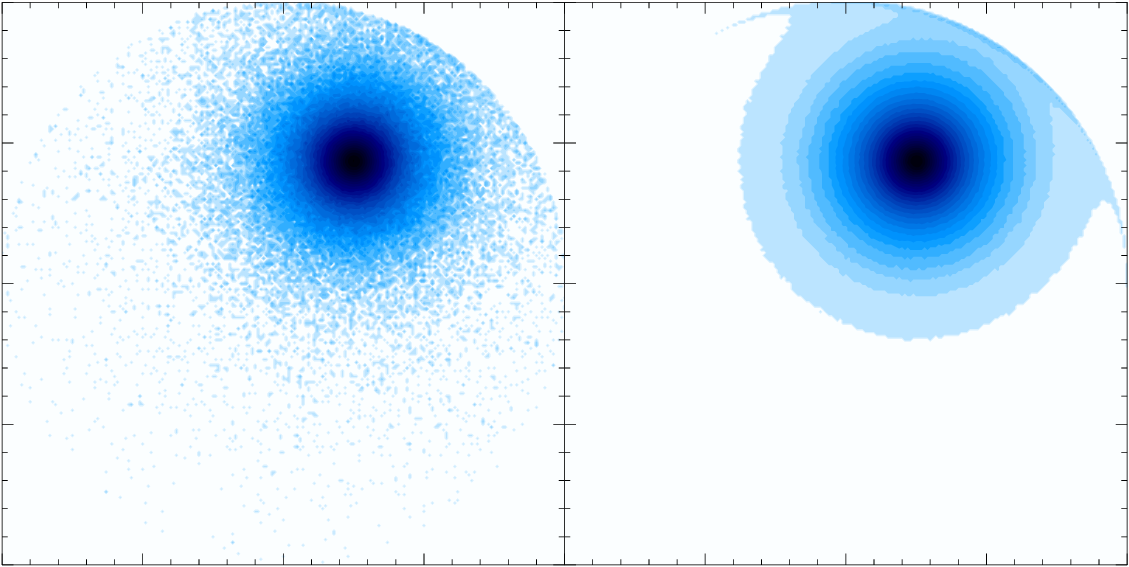}
\end{center}
\caption{\label{fig:label3} Comparison of the Monte Carlo implementation (left) and the analytic solution of the Harvey-Shack scattering for an ABC surface (right), shown on a logarithmic intensity scale.  We choose $B=10^{-2}~\mbox{mm}$ and $C=3.5$ to emphasize a larger scattering angle than a typical polished surface to optimally compare the approaches. }
\end{figure}

Then we reject solutions with random probability less than the value, $F$.

\begin{equation}
  F = \frac{ \left( \cos{\theta} + \cos{\theta_i} \right)^2 }{ \left( 1 + \cos{\theta_i} \right)^2 } \frac{ f_c \cos{\theta_i}}{\cos{\theta}}
  \label{equation25}
  \end{equation}

\noindent
where we choose a numerical constant, $f_c=0.9$, when $F$ is very unlikely to exceed 1.  If the test fails, we repeat the scattering beginning with a new value of $\alpha^{\prime}$.  We have tested this procedure to determine the accuracy of the Monte Carlo approach with the analytic prediction of the two dimensional distribution.  In Figure~\ref{fig:label3}, we simulate scattering from a surface roughness pattern having $B=10^{-2}~\mbox{mm}$ and $C=3.5$.  The overall scattering roughness parameter, $\sigma$, is unimportant since we have confirmed the normalization matches exactly. The overall pattern, central scattering core, and even the scattering at grazing angles is reproduced.  Any systematic mismatch is below the statistical noise with the 1 million simulated photons and has any systematic error less than $0.5\%$.  The above can be implemented in several lines of code and has minimal mathematical runtime calculations.

\begin{figure}[htb]
\begin{center}
\includegraphics[width=0.99\columnwidth]{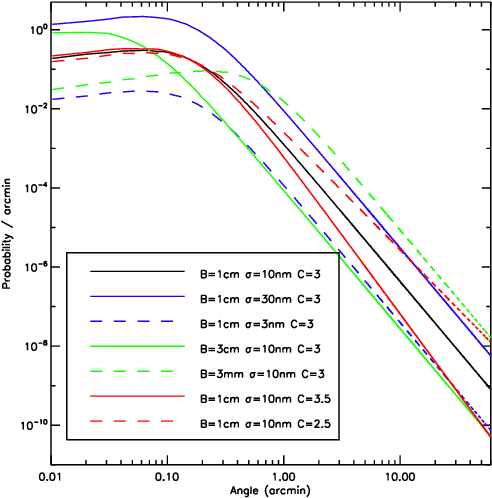}
\end{center}
\caption{\label{fig:label4} The angular scattering profile based on several roughness parameter choices.  The black line shows $B=1~\mbox{cm}$, $\sigma=10~\mbox{nm}$, and $C=3$.  The blue lines increase (solid) or decrease (dashed) the normalization to $\sigma=30~\mbox{nm}$ and $\sigma=3~\mbox{nm}$, respectively.  The green lines increase (solid) or decrease (dashed) the value of $B=3~\mbox{cm}$ and $B=3~\mbox{mm}$.  The red lines increase (solid) or decrease (dashed) the value of $C=3.5$ and $C=2.5$.  Therefore, $\sigma$ controls the total amount of scatter, $B$ sets the angular scale, and $C$ determines the roll off at large angle.}
\end{figure}

Based on this representation, we can describe an arbitary surface by the three parameters:  $\sigma$, $B$, and $C$.  $\sigma$ sets the overall normalization and thus describes how rough the surface is.  $B$ represents the spatial scale where the spectrum steepens.  Since the scattering to larger angles corresponds to smaller spatial patterns, then increasing $b$ results in less angular scattering.  It is likely, the process of polishing a surface results in both decreased total roughness, $\sigma$, as well as increasing the value of $B$.  Finally, the parameter $C$ sets the angular fall off of the scattering pattern and is often around 3, and likely a material fractal property of surfaces.  This  behavior is shown in Figure~\ref{fig:label4} where we calculate various profiles based different values of the three parameters.

\subsection{Comparison with Data}

To confirm the representation above, we can compare to real point spread function wings.   We compare with observations taken with the WIYN telescope (\citealt{johns1994}, \citealt{johns1990}).  It has a 3.5m primary borosilicate mirror (\citealt{angel1984}) that is actively controlled (\citealt{roddier1995}).  We use observations using the One Degree Imager (ODI) camera (\citealt{jacoby2002}).

We have collected hundreds of one minute exposures using WIYN ODI in a variety of filters with fields containing galaxy clusters.  To study the large angle scattering we selected exposures in the g, r, and i filters.  We then obtained cutout images of isolated bright stars that were unsaturated using a pipeline described in \cite{dutta2024}.  All 146 \textit{i} band, 151 \textit{r} band, and 161 \textit{g} band images were co-added using the SWARP.   Sources in the coadded image were detected with Sextractor.  The SWARP and Sextractor parameters used can be found in \cite{dutta2024}. The sources are cross-matched with Gaia EDR3 (\citealt{gaia2016}, \citealt{gaia2020}). Any source within 0.72” of the Gaia catalog is considered a star. If there are multiple sources in that radius, the brightest one is designated as the star. All stars where {\it bflag} or {\it vflag} are raised are rejected. The flags check for any unusal pixel values or the vertical line of nan’s though extremely bright stars.   When making cutouts of stars in individual exposure, we reject any star brighter than $10^9$ counts. To reject blended sources we find k=3 sigma clipped median and standard deviation of $\sigma_{xx}$ and $\sigma_{yy}$ (the second moments in the x and y directions). For individual exposures, stars where flags are raised are also rejected. All cutouts are 100x100 pixel in size.

The ability to determine the origin of certain PSF components is complex.  The profile predicted in Figure~\ref{fig:label4} will be most visible in the wing of the PSF profile.  Since it falls off rapidly with radius, we would predict between 1'' and 3'' from the center before it blends in with the background.  The core of the PSF for WIYN is dominated by the atmospheric turbulence.  Turbulence PSF profiles and patterns have been extensively studied  (\citealt{lane}; \citealt{trujillo2001}; \citealt{ellerbroek}; \citealt{lelouarn}; \citealt{britton}; \citealt{jolissaint}) and has been extensively modelled in PhoSim (\cite{chang2012}, \cite{peterson2015}).  There will also be a contribution from the optical aberrations and deformations on the surface of the mirror and the charge diffusion in the detector.  The optics deformations of WIYN are smaller than a pixel (\citealt{roddier1995}) and estimated by our simulations \citep{peterson2019}.  This should have a negligible contribution to the PSF wing.  Similarly, the charge diffusion produces a quasi-gaussian blur which is also below a pixel size and has no wing for the 30 micron thick devices of ODI (\citealt{peterson2020}).  Another possible contribution to the PSF wing is the scattering from aerosols including water vapor (\citealt{devore2013}).
However, this is much more significant for very bright stars where the profile can be studied to few arcminute scales.   On arcsecond scales this should be negligible because the profile is effectively flat, except in the cases of significant cloud cover (\citealt{peterson2024}).   Therefore, our general strategy is that the PSF wing between 1'' and 3'' should be dominated by a combination of a residual wing from the turbulence pattern and mirror microroughness.  Furthermore, using fairly clear nights with better seeing should make the microroughness more visible.

For the fields discussed above, we obtained between 700 and 1100 stars.  We removed exposures where the background was higher than 150 counts per pixel in order to see the PSF wing better.  We also removed exposures where the PSF size was bigger than 1'' (FWHM).  Furthermore, we removed exposures where the PSF ellipticity was larger than 0.4 due to telescope tracking errors.  We then had a total of 60 exposures of stars in the g, r, and i bands.   For each exposure, we obtained the radial profile of counts and added the counts from the roughly one thousand stars to have a single PSF radial profile for each exposure.  We then fit the profile with three components.  First, we add a flat background component.  Second, we added a profile representing the PSF due to turbulence.  This profile is not simple, since turbulence itself in the long exposure limit produces a quasi-gaussian PSF with a wing.  The wing is due to high spatial frequency fluctuations in the index of refraction.  To study this, we simulated PSFs using PhoSim using purely turbulence.  We found that the profile can be fit to the same form as the microroughness scattering profile

\begin{equation}
  P=\frac{1}{{\left[ 1+{\left( \frac{r}{b} \right)}^2\right] }^{\frac{c+1}{2}}}
  \label{equation26}
\end{equation}

\noindent
where $c = 3.7$, $b = 1.15 s_0$, and $s_0$ is the seeing.   Since it has a similar form, it makes the fitting somewhat degenerate and is the reason for carefully only choosing exposures where the profile will be well-measured.  Fortunately, it has a much steeper profile (due to the Kolmogorov nature of the turbulence), so we can still measure the microroughness wing at large angle.  We ignored the full simulations of the optics and detector since it is different for every star and would be negligible in studying the wing as discussed above.  It could slightly bias the seeing estimate, but will not affect the microroughness analysis.
Finally, we add a third component to represent the microroughness.  The microroughness profile is convolved with the turbulence profile using a Monte Carlo to predict the PSF of the stellar photons.  Then we have a total of 5 free parameters (the background level, the seeing, and the three parameters of the ABC model).   It is possible that there is also some component of the wing due to Mie scattering in the atmosphere off of water and other molecules as predicted in \cite{peterson2024}, but we found this is only important when there is significant cloud  absorption that we excluded from our sample.

\begin{figure}[htb]
\begin{center}
\includegraphics[width=0.99\columnwidth]{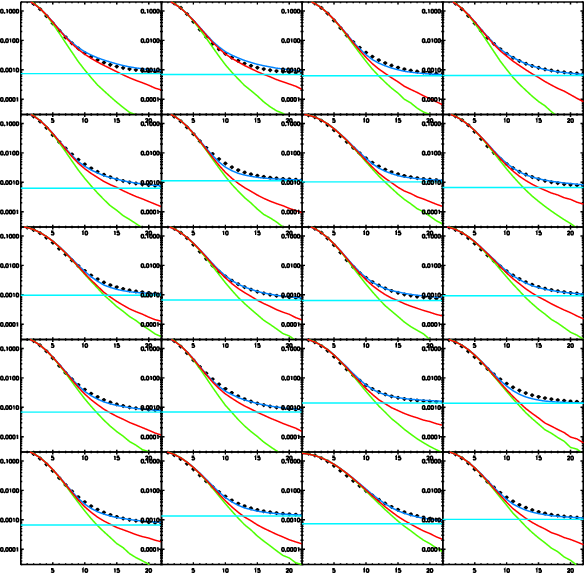}
\end{center}
\caption{\label{fig:label5} Radial co-added PSF profiles (fractional flux vs. pixel) of WIYN ODI observations of bright stars.  The pixel size is 0.11 arcseconds.   In each panel, 700 to 1000 star PSF profiles are added together, and then fit using the model described in the text.  The light blue is the background, the green is the turbulence profile, and the red convolves the turbulence profile with the microroughness scattering profile.  The total of all three is the dark blue model that should match the data.}
\end{figure}

\begin{figure}[htb]
\begin{center}
\includegraphics[width=0.9\columnwidth]{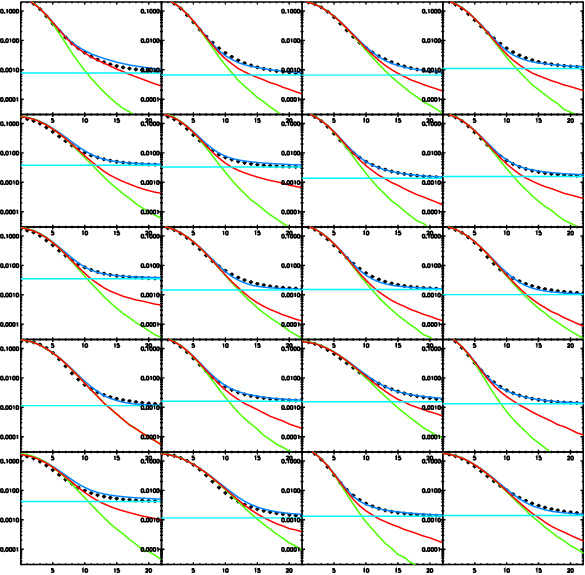}
\end{center}
\caption{\label{fig:label6} Radial co-added PSF profiles as described in Figure~\ref{fig:label5}.}
\end{figure}

\begin{figure}[htb]
\begin{center}
\includegraphics[width=0.9\columnwidth]{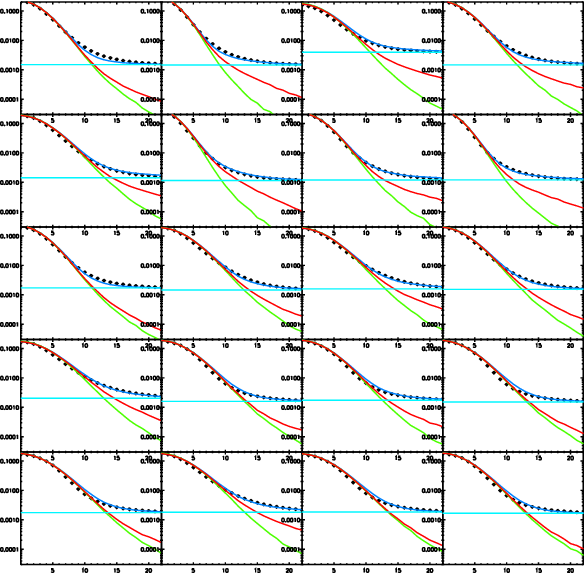}
\end{center}
\caption{\label{fig:label7} Radial co-added PSF profiles as described in Figure~\ref{fig:label5}.}
\end{figure}

\begin{figure}[htb]
\begin{center}
\includegraphics[width=0.9\columnwidth]{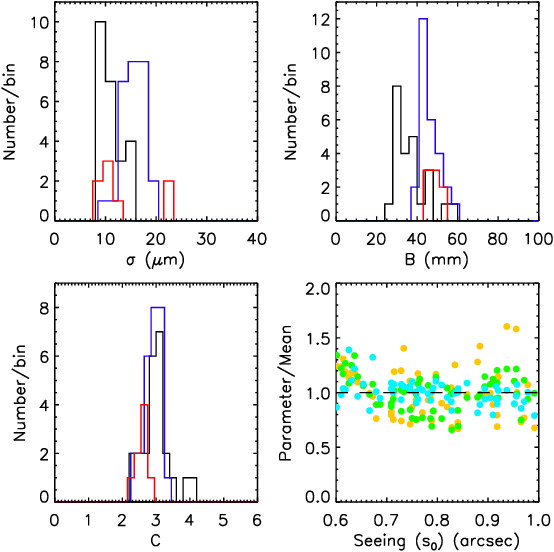}
\end{center}
\caption{\label{fig:label8}  The first three panels show histograms of the fitted parameters of the microroughness model described in the text.  The g band observations are shown in black, the \textit{r} band observations are shown in blue, and the i band observations are shown in red.  The bottom right panel shows the lack of correlation between the three microroughness parameters ($\sigma$=orange, $B$=green, and $C$=cyan) and the seeing of the core of the PSF.}
\end{figure}

The profiles and the result of the fits is shown in Figures~\ref{fig:label5},~\ref{fig:label6}, and~\ref{fig:label7}.  We use an MCMC method and obtain a stable chain of at least 1000 iterations.  In Figure~\ref{fig:label8} we show the results of the fits.  We obtain for these exposures a consistent set of parameters describing the microroughness of the mirror.  First, we obtain a microroughness normalization of $15 \pm 3$ nm, which implies a scattering fraction of 0.12 at 500 nm.  Second, we measure $B = 46 \pm 8$ mm.  This is a reasonable spatial scale where it would be difficult to polish.  Third, we measure $C = 3.1 \pm 0.3$.  These parameters seem reasonably consistent between our observations.  There is also limited differences between bands, and no obvious trend between different bands.   This is important, since we included the wavelength-dependence according to the representation, then both $\sigma$ and $B$ should have wavelength-independent values.

 We also investigated any correlation with the microroughness parameters with the seeing in the bottom right panel of Figure~\ref{fig:label8}.  Seeing is both highly variable in time and varies by a factor of 1.7 in our sample.  There is no obvious correlation and any degeneracy between the seeing parameter and the microroughness parameters could at most distort the parameters by 20$\%$.   Thus, we can represent the microroughness profile pattern by a single set of parameters.  Within the error range, this analysis implies that we have correctly separated the static surface microroughness pattern from the time-variable atmosphere and cloud profile.

 Note that for determining roughness parameters for other telescopes, since we established the validity of this approach, we can simply forward model a set of three parameters to match the observed PSF wing profiles.  In this case, we do not need to isolate the bright stars that are not unsaturated, remove the bad seeing, and carefully fit the profiles.  Instead, the three parameters can be modified with a single line for each mirror in the PhoSim instrument characteristics and the whole image will acquire the PSF wing blur that is generally quite easy to observe.  Even the saturated stars could more accurately determine the profile.  Figure~\ref{fig:label4} demonstrates how the three parameters are not coupled and affect the normalization, the shape several arcseconds from the core, and the large scale roll off.

\section{Contamination}

Contamination on surfaces in the form of both dust and condensation can alter the trajectory of light.   To properly describe the photon interaction, we can apply Mie-Tyndall theory (\citealt{tyndall1869}, \citealt{mie1908}) where it produces a self-consistent electromagnetic calculation.   This can then predict the absorption and scattering probability as a function of angle.  Mie-Tyndall scattering depends  on the dimensionless ratio, $ x = 2 \pi r / \lambda$, where $r$ is the size of the particle and $\lambda$ is the photon wavelength, as well as the real and imaginary part of the index of refraction.  We use an identical approach to \cite{peterson2024} and produce a grid of values as a function of $x$ and $\lambda$ using the Bohren-Huffman algorithm (\citealt{bohren1983}).  This solves the electromagnetic boundary conditions in an iterative approach, and we obtain the angular scattering probability distribution and absorption probability.

For contamination, we therefore represent the dust and condensation as spherical uniform objects.   We assume that the size of each component of the contaminants follows a log normal distribution.  Therefore for a distribution with mean size, $\mu$, in microns and standard deviation,  $d$, the surface density, $c_s$, in particles per square microns is given by

\begin{equation}
   c_s = \frac{c_f}{ \pi e^{2 \log{\mu} + 2 d^2}}
  \label{equation27}
  \end{equation}

\noindent
where the fraction of the surface that is covered by contaminants is $c_f$.  We have squared the mean of the log-normal distribution to calculate the denominator, and estimate the average projected area of the contaminants.  It is likely that $c_f$ is near a few percent, but this can be studied with various surfaces.  We then add dust particles across an optical surface where the total number of particles is given by $c_s A$, where $A$ is the area of the surface.  For the dust we use the optical constants of \cite{aspnes1983} and assume $\mu = 10~\mu\mbox{m}$ and  $d = 1.0$.  For the condensation we use the optical constants of \cite{hale1973} and assume $\mu = 20~\mu\mbox{m}$ and $d = 1.0$.

The size distribution of dust or contamination for mirror surfaces is not well known from the literature.   This is somewhat complicated because the size distribution of contaminants in the atmosphere is not the same (and is smaller) than the size distribution of what ends up on surfaces.  Contamination on surfaces has been studied in clean room environments (\citealt{hamberg1984}, \citealt{mil1994}).   Log-normal distributions are assumed and the larger contaminants can be measured.  The large particle distribution  is often compared with a linear relation between the logarithmic particle density per area above a given size and the square of the logarithm of particle size.   \cite{hamberg1984} find an empirical slope of 0.383, whereas \cite{mil1994} find a slope of 0.926.  Either of these can be reproduced with a log-normal distribution by varying the mean and  standard deviation in the representation we are using above.
However, the normalization, $c_f$, is hard to predict from these ideal environments, because telescope mirrors are exposed to the environment at various angles, with fairly laminar air flows, and can be cleaned periodically.  A somewhat similar application is the investigation by \cite{zucker2008} for LIGO mirrors, but with obviously a different environment.   In future work, we can determine the normalization and match the distribution in more detail, but it is likely that $c_f$ is in the 0.1$\%$ to 10$\%$ percent range for typical telescopes.  Those levels would be enough to see the evidence dust contamination, but not dominate the photometric response.

To perform the computation, we first divide the surface into a Cartesian grid having typically 256$\times$256 elements.  Then we loop through the total numbers of particles.
To simulate the potentially prohibitive number of contaminants, we use three regimes based on the size of the contaminant.   We choose particles according to the log-normal distribution. For the largest of contaminants, we select particles with a size larger than $\epsilon G$, where $G$ is the Cartesian grid size and $\epsilon$ is a numerical constant nominally set to 0.3.  For these contaminants, we simulate the entire interaction using Mie-Tyndall theory in an identical approach to aerosol interactions in \cite{peterson2024}.  Here we calculate the value of $x = 2 \pi c_r / \lambda$, where $c_r$ is the contaminate size.   We randomly choose the polar scattering angles using the Mie-Tyndall distribution.  We also potentially lose some photons according the predictive absorption probability.  To keep track of these interactions, we have to keep a list of at least two particles for every cell, so that when we simulate the light hitting the surface we will be able to quickly evaluate if it is closest enough to consider an interaction.  Then, we potentially only have a few large contaminants to actually perform a Mie scattering calculation for each photon.

For moderate size contaminants that are smaller than $\epsilon G$ but larger than $\epsilon^2 G$, we simulate the interaction by assuming that we lose a fraction of light in each grid cell according to the total accumulated contamination area in that grid cell.  This will then still capture the uneven nature of the dust pattern, because there will be some Poisson fluctuation in each grid cell.  However, these particles are small enough that they will not result in observable features that the larger contaminants cause.  Finally for the smallest contaminants where their size is less than $\epsilon^2 G$, we simply keep track of the total area across the entire surface and reduce their transmission by a constant factor.  The combined approach of this method is we can potentially consider the interaction of billions of contaminants with every single photon, with a straight-foward look up of the location of the large contaminants and a transmission fraction calculation for each grid cell.

\subsection{Effect of Contamination}

The contamination can lead to dust rings where certain angles are blocked due to the presence of the contaminant.  This depends on the distance, $L$, from the focal plane to the contaminated surface.  The dust ring will have diameter of $L / \mathcal{F}$, where $\mathcal{F}$ is the f-number of the optical system.  A fraction of the background will be reduced by an amount equal to ${c_d^2} \mathcal{F}^2 / {L^2} $, where $c_d$ is the diameter of the contaminant.  In addition, the total amount of light lost is $ \pi c_d^2 / \left( 4 \mathcal{F}^2 \right)$.  In many modern telescopes, $L$ can be chosen to be large enough so dust rings are of order the size of the sensor and therefore cause less of a prominent pattern.  However, the light is still lost and the overall photometric pattern will need to be corrected.  In the next section, we simulate some dust patterns with $L = 1$ mm to exaggerate the effect.  Note that the relations above assume that $c_d$ is small compared to $L$, but if that is not the case then $L$ should be replaced with $\sqrt{ L^2 + c_d^2}$ in the first two expressions.   This also leads to a more gradual rim having size of $c_d / \mathcal{F}$.  Furthermore, it appears that the sharpness of the rim is slightly affected by the Mie-Tyndall distribution, so it may be possible to learn the nature of the contaminant.

\section{Simulation of Uncoated, Unpolished \& Contaminated Mirror}

\begin{figure}[htb]
\begin{center}
\includegraphics[width=0.89\columnwidth]{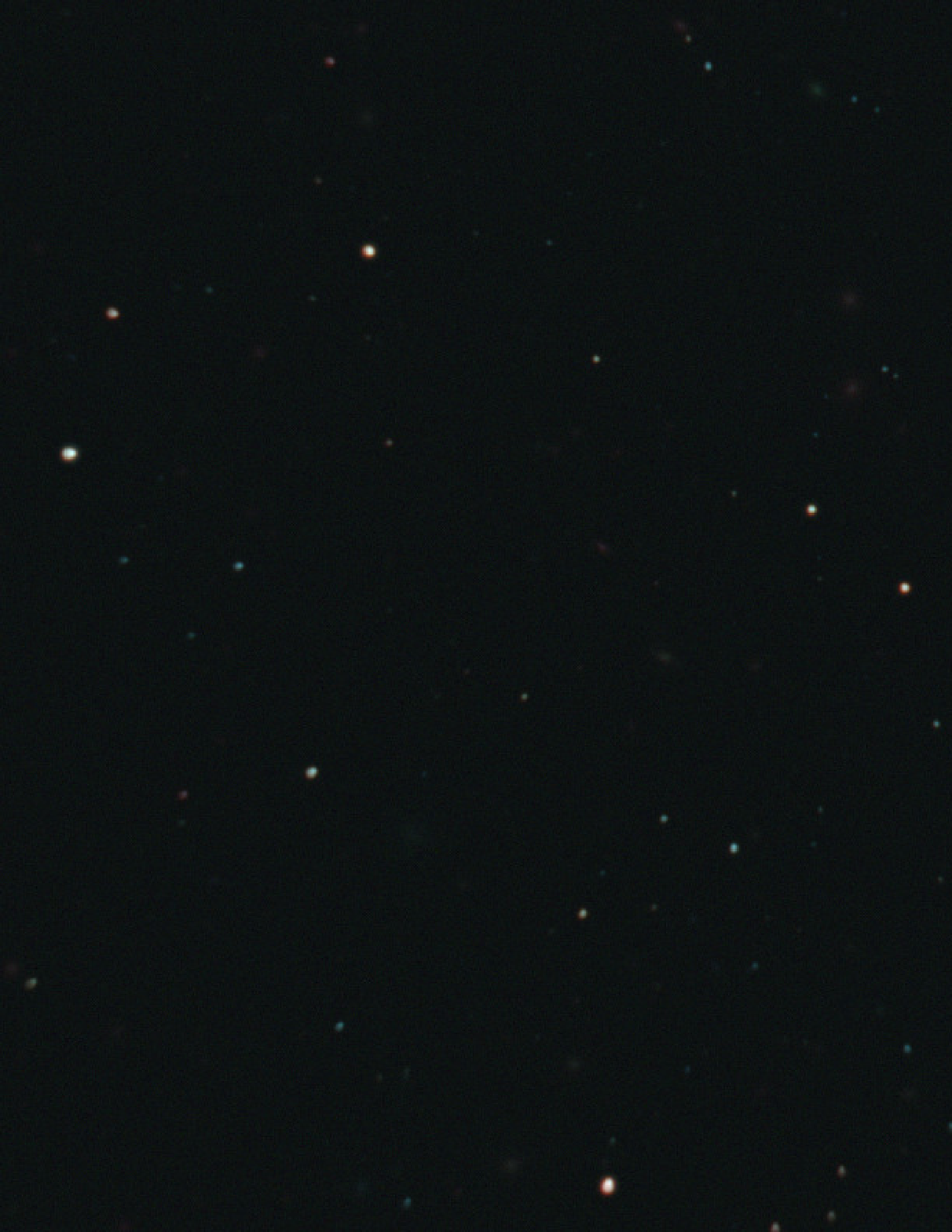}
\end{center}
\caption{\label{fig:label9} A simulation of a generic telescope with a Silver mirror coated with Silicon Nitride, and a lens surface with a Magnesium Fluoride coating.  The mirror has no microroughness and there is no contamination on any of the surfaces.  A set of stars and galaxies is seen, and this image can be compared with Figure~\ref{fig:label10}.  The rgb colors are set to the U, V, and I simulations with a logarithmic scale, and the field size is approximately 6 by 7 arcminutes.}
\end{figure}

\begin{figure}[htb]
\begin{center}
\includegraphics[width=0.89\columnwidth]{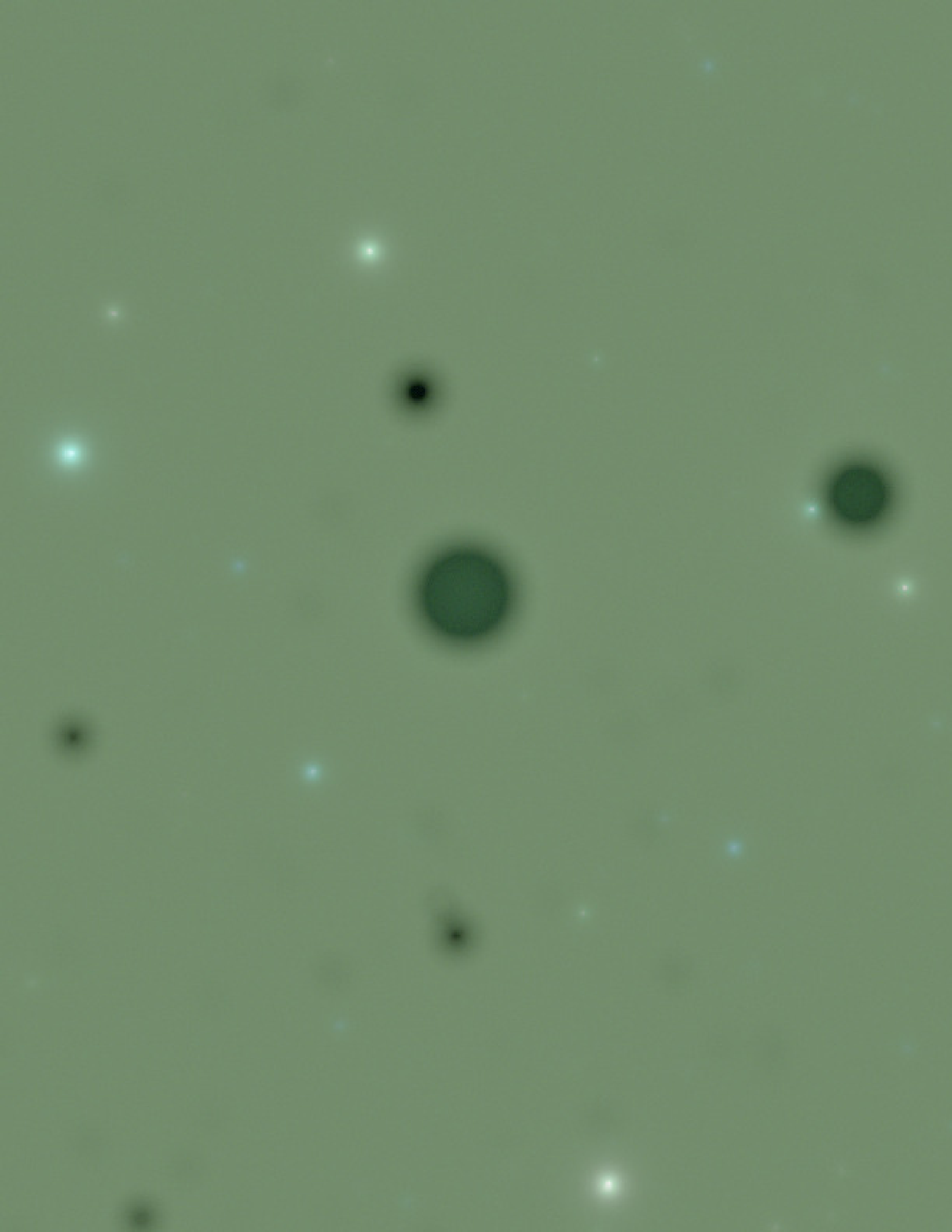}
\end{center}
\caption{\label{fig:label10} The same simulation as Figure~\ref{fig:label9} but with no coatings on the surfaces, significant microroughness on the mirror, and significant contamination of the lens surface.  The difference in colors due to the coatings, the large PSF due to the microroughness, and the loss of light due to the contamination is clearly visible.  Thus, the major elements of surface interaction are exaggerated in this image.}
\end{figure}

The combination of all three elements of surface interaction can be shown in a single simulation.   We first simulate a 1 m generic parabolic telescope with a 4 m focal length with a 1k $\times$ 1k sensor with 10 micron pixels.  We use the default star and galaxy synthetic catalogs as described in \cite{burke2019a}, and simulate the objects in 3 filters corresponding with traditional U, V, and I filters.  This simulation is shown in Figure~\ref{fig:label9} where there is no microroughness on the mirror, a Silver mirror with 0.083 $\mu m$ Silicon Nitride coating, the second surface of the filter with a 0.18 $\mu m$ Magnesium Fluoride coating, and no contamination on any surface.

In Figure~\ref{fig:label10}, we add the three major elements studied in this paper.  First, we remove the two coatings on the surfaces which affects the total throughput.   Second, we add 1\% of surface contamination split equally between dust and condensation on the second surface of the filter.  This surface is located 1 mm from the sensor, which then makes the dust rings more prominent.  In general, we can put different amounts of dust and condensation on any surface, but we simulate this to emphasize the dust rings.   Finally, we add significant microroughness on the surface of the mirror to show significant scattering.  We set the microroughness with $\sigma = 100~\mbox{nm}$, $B = 50~\mbox{mm}$ and $C = 2$ to make the wings more apparent.  As can be seen in Figure~\ref{fig:label10} there is a significant PSF for the same stars in Figure~\ref{fig:label9}.  The different colors are also evident due to the effect of the coatings.  The dust rings are also easily seen.\footnote{In PhoSim, use a mirror coating file with {\it protectedlayer si3n4 ag 0.083} for Figure~\ref{fig:label9} and a mirror coating file with {\it scattering 100e-9 0.05 2.0} and a lens coating file with {\it contamination 0.005 0.005} for Figure~\ref{fig:label10}.}

\section{Conclusion \& Future Work}

We have presented a comprehensive Monte Carlo implementation of the major elements of surface interaction including the interference with surfaces including coatings, the effect of scattering due to surface microroughness, and the complex interaction with contaminants on surfaces.   The implementation add little computation cost when embedded in a photon Monte Carlo even for very complex surfaces, microroughness profiles, and  significant contamination.  We have disentangled the PSF patterns from turbulence from mirror microroughness with an \textit{ab initio} approach to each.  The methods are comprehensively implemented in PhoSim v6.1.  This version is publicly available (\url{https://www.phosim.org}).

Future work will be done with the community to validate aspects of this implementation with a variety of telescopes.  The contamination levels can be studied with real telescopes to validate the particle size distributions and overall transmission normalization.  Similarly, the microroughness profiles can be studied by comparing with other telescopes and studying the PSF wings.  Finally, the transmission and reflection of surfaces including complex coatings can be compared with throughput measurements.  The non-uniformity of coatings in particular can also be studied to understand one of the most important contributions to photometric error.

We also expect that the formalism in this paper will also be useful for non-astronomical applications.  In particular, the microroughness scattering, the contamination, and surface throughput formalism would apply to the interaction of light with everyday objects as well.  Basically, how we see objects is dictated by the physics in this paper.  By implementing the physics in a very efficient photon Monte Carlo approach, it could be useful for understanding of the perception of objects and for simulating objects in computer graphics applications.

\begin{acknowledgments}

We thank Purdue University for support and an anonymous referee for helpful comments.

\end{acknowledgments}

\end{document}